\documentclass[12pt]{article}

\def\fft#1#2{{#1\over#2}}

\def\hoch#1{$\, ^{#1}$}

\begin{document}


\begin{flushright}
UM-TH-00-27\\
hep-th/0010171\\
\end{flushright}

\vspace{40pt}

\begin{center}

{\bf Hodge Duality on the Brane}

\vspace{20pt}

M.J. Duff\hoch{1} and
James T. Liu\hoch{1}

\vspace{10pt}
{\it
Randall Laboratory, Department of Physics, University of Michigan,\\
Ann Arbor, MI 48109--1120}

\vspace{30pt}

\underline{ABSTRACT}
\end{center}
\noindent
It has been claimed that whereas scalars can be bound to a
Randall-Sundrum brane, higher $p$-form potentials cannot, in
contradiction with the Hodge duality between $0$-form and
$3$-form potentials in the five-dimensional bulk. Here we show
that a $3$-form in the bulk correctly yields a $2$-form on the
brane, in complete agreement with both bulk and brane duality.
We also emphasize that the phenomenon of photon screening in the
Randall-Sundrum geometry is ruled out by the bulk Einstein equation.

{\vfill\leftline{}\vfill
\vskip 10pt \footnoterule
{\footnotesize\noindent\hoch{1}
Research supported in part by DOE Grant DE-FG02-95ER40899 Task G.
\vskip  -12pt} \vskip   14pt
}

\newpage


It is well-known that in the Randall-Sundrum brane world picture
\cite{Randall}, a massless scalar field in the $(d+1)$-dimensional
bulk results in a massless scalar being bound to the $d$-dimensional
brane.  In recent work \cite{kaloper}, however, it is claimed
that massless $p$-form fields with $p \geq (d-2)/2$ in the bulk yield no
massless fields on the brane. In particular, only scalars are bound
to the brane in $d=4$. The argument runs as follows. The action for
a massless $p$-form potential $\hat A_{[p]}$ with $(p+1)$-form field
strength $\hat F_{[p+1]}=d \hat A_{[p]}$ in the $(d+1)$-dimensional
bulk is given by
\begin{equation}
S_{\rm bulk}=\int d^{d+1}x
\left[-\frac{1}{2(p+1)!}\sqrt{-\hat g}\hat g^{M_{1}N_{1}}
\hat g^{M_{2}N_{2}}\cdots \hat g^{M_{p+1}N_{p+1}}
\hat F_{M_{1}M_{2}\ldots M_{p+1}}
\hat F_{N_{1}N_{2}\ldots N_{p+1}}\right].
\label{bulkaction}
\end{equation}
Denoting the coordinates $x^{M}=(x^{\mu},z)$, the Kaluza-Klein ansatz
for the metric is
\begin{equation}
\hat g_{MN}dx^{M}dx^{N}=e^{-2k|z|}g_{\mu\nu}(x)dx^{\mu}dx^{\nu}+dz^{2}.
\label{metric}
\end{equation}
If the $p$-form ansatz is
\begin{equation}
\hat A_{\mu_{1}\mu_{2}\ldots\mu_{p}}(x,z)
=A_{\mu_{1}\mu_{2}\ldots\mu_{p}}(x),
\label{pform}
\end{equation}
with all other components vanishing (so that the field
strength reduces straightforwardly, $\hat F_{[p+1]} = F_{[p+1]}$),
then the resulting brane action is
\begin{eqnarray}
S_{\rm brane}&=&\int d^{d}x
\left[-\frac{1}{2(p+1)!}\sqrt{-g}g^{\mu_{1}\nu_{1}}g^{\mu_{2}\nu_{2}}\cdots
g^{\mu_{p+1}\nu_{p+1}} F_{\mu_{1}\mu_{2}\ldots \mu_{p+1}}
F_{\nu_{1} \nu_{2} \ldots \nu_{p+1}}\right]\nonumber\\
&&\kern8em\times \int dz\, e^{2\left(p+1-\frac{d}{2}\right) k|z|}.
\label{braneaction}
\end{eqnarray}
The criterion for being bound to the brane is the convergence of the
$z$ integral, or
\begin{equation}
p<\frac{d-2}{2}.
\label{converge}
\end{equation}
So the ansatz (\ref{pform}) would imply in particular that only $0$-forms
can be bound to the brane in $d=4$.

However, this argument is in contradiction with the well-known result
that, in the absence of topological obstructions \cite{DuffvanN}, a
$p$-form potential $\hat A_{[p]}$ in the bulk is dual to a $(d-p-1)$-form
potential $\kern4pt\hat{\kern-4pt\tilde A}_{[d-p-1]}$ with $(d-p)$-form
field strength $\kern2pt\hat{\kern-2pt\tilde F}_{[d-p]}=d
\kern4pt\hat{\kern-4pt\tilde A}_{[d-p-1]}$:
\begin{equation}
\kern-2pt
\sqrt{-\hat g} \hat g^{M_{1}N_{1}} \hat g^{M_{2}N_{2}}\cdots
\hat g^{M_{d-p}N_{d-p}}
\kern2pt\hat{\kern-2pt\tilde F}_{N_{1}N_{2}\ldots N_{d-p}}=
\frac{1}{(p+1)!}\epsilon^{M_{1}M_{2}\ldots M_{d-p} N_{1}N_{2}\ldots N_{p+1}}
\hat F_{N_{1}N_{2}\ldots N_{p+1}}.
\kern-3pt
\label{bulkdual}
\end{equation}
Indeed, it is easy to see that the Kaluza-Klein ansatz (\ref{pform})
cannot be chosen for both $\hat A_{[p]}$ and
$\kern4pt\hat{\kern-4pt\tilde A}_{[d-p-1]}$
simultaneously since it would not be compatible with (\ref{bulkdual}).

The resolution of the paradox is to keep the ansatz (\ref{pform}) for
$p<(d-2)/2$ but modify it for $p \geq (d-2)/2$. First, for $p>d/2$ we write
\begin{equation}
\hat A_{\mu_{1}\mu_{2}\ldots \mu_{p-1} z}(x,z)=
e^{-2\left(p-\frac{d}{2}\right) k|z|}A_{\mu_{1}\mu_{2}\ldots\mu_{p-1}}(x),
\label{newpform}
\end{equation}
with other components vanishing, so that
\begin{equation}
\hat F_{\mu_{1}\mu_{2}\ldots \mu_{p} z}=e^{-2\left(p-\frac{d}{2}\right)k|z|}
F_{\mu_{1}\mu_{2}\ldots \mu_{p}}.
\end{equation}
This is now perfectly compatible with (\ref{bulkdual}). One finds
\begin{equation}
\sqrt{-g} g^{\mu_{1}\nu_{1}} g^{\mu_{2}\nu_{2}}\cdots g^{\mu_{d-p}\nu_{d-p}}
{\tilde F}_{\nu_{1}\nu_{2} \ldots \nu_{d-p}}=
\frac{1}{p!}\epsilon^{\mu_{1}\mu_{2} \ldots \mu_{d-p} \nu_{1}\nu_{2}
\ldots \nu_{p} }
F_{\nu_{1} \nu_{2} \ldots \nu_{p}},
\label{branedual}
\end{equation}
where $F_{[p]}$ and $\tilde F_{[d-p]}$ result from ansatze (\ref{pform})
and (\ref{newpform}) respectively.
So duality in the bulk implies duality on the brane as it must for
consistency. The brane action now describes a $(p-1)$-form potential:
\begin{equation}
S_{brane}=\int d^{d}x
\left[-\frac{1}{2\cdot p!}\sqrt{-g}g^{\mu_{1}\nu_{1}}g^{\mu_{2}\nu_{2}}\ldots
g^{\mu_{p}\nu_{p}}
F_{\mu_{1}\mu_{2}\ldots \mu_{p}}F_{\nu_{1}\nu_{2} \ldots \nu_{p}}\right]
\int dz\, e^{-2\left(p-\frac{d}{2}\right) k|z|}.
\label{newbraneaction}
\end{equation}
The criterion for being bound to the brane, or convergence of the $z$
integral, is now
\begin{equation}
p>\frac{d}{2}.
\label{newconverge}
\end{equation}
In particular for $d=4$, a $p$-form with $p>2$ in the bulk yields a
$(p-1)$-form on the brane. This now resolves the original
duality paradox. Thus a $3$-form which is dual to a scalar in
the five-dimensional bulk yields a $2$-form which is dual to a scalar
on the four-dimensional brane.

For the intermediate values $p=d/2$ and $p=(d-2)/2$, neither the
$p$-form nor the $(p-1)$-form is bound to the brane. Note in particular
therefore that photons are not bound to the brane in $d=4$ if they
are described by Maxwell's equations in the bulk. In this respect, we
agree with \cite{Bajc:2000mh,kaloper}%
\footnote{In the supersymmetric Randall-Sundrum brane world
\cite{Bremer,Duffliu,DLS}, the graviphotons on the $d=4$ brane have a
different origin in terms of odd-dimensional self-duality equations
in the bulk \cite{lupope,Duffliu2,Cvetic:2000gj}. Similarly in $d>4$, higher
$p$-forms may be bound if their bulk origin is not given by
(\ref{bulkaction}).}

A few comments on the Kaluza-Klein reduction are now in order.  Firstly, the
ansatze (\ref{pform}) and (\ref{newpform}) are not arbitrary, but must be
chosen to respect the bulk $p$-form equation of motion,
\begin{eqnarray}
\label{eq:eom1}
\hat \nabla^{M_1}\hat F_{M_1M_2\ldots M_{p+1}}=0.
\end{eqnarray}
In the Randall-Sundrum background, (\ref{metric}), this becomes
\begin{eqnarray}
\label{eq:eom2}
&&\hat \nabla^{\mu_1}\hat F_{\mu_1\mu_2\ldots\mu_{p+1}}
+e^{-2\left(p+1-\fft{d}2\right)k|z|}\partial_z
\left(e^{2\left(p-\fft{d}2\right)k|z|}
\hat F_{z\mu_2\mu_3\ldots\mu_{p+1}}\right)=0,\nonumber\\
&&\hat \nabla^{\mu_1}\hat F_{\mu_1\mu_2\ldots\mu_pz}=0.
\end{eqnarray}
Kaluza-Klein reduction of a $p$-form potential generically yields both
$p$-form and $(p-1)$-form potentials on the brane.  A natural ansatz in the
warped background would be to choose
\begin{equation}
\label{eq:aantz}
\hat A_{[p]}(x,z)= A_{[p]}(x) + A_{[p-1]}(x) \wedge df(z),
\end{equation}
where $f(z)$ is {\it a priori} an arbitrary function of $z$.  Note
that there is a freedom of gauge choice in making the ansatz
(\ref{eq:aantz}).  One may transform to a gauge
$\hat A_{\mu_1\mu_2\ldots\mu_{p-1}z}=0$, whereupon (\ref{eq:aantz}) becomes
\begin{equation}
\hat A_{[p]}(x,z)=A_{[p]}(x)+(-1)^pf(z)F_{[p]}(x).
\end{equation}
This is the form of the ansatz used in \cite{lupope,Cvetic:2000gj}.
Either way, substituting
the resulting field strength
\begin{equation}
\hat F_{[p+1]}(x,z)=F_{[p+1]}(x)+F_{[p]}(x)\wedge df(z)
\end{equation}
into (\ref{eq:eom2}), one then obtains the $d$-dimensional equations of
motion,
\begin{equation}
d*_dF_{[p+1]}=0,\qquad d*_dF_{[p]}=0,
\end{equation}
along with the constraint $f'(z)=e^{-2\left(p-\fft{d}2\right)k|z|}$.  This is
the origin of the ansatze (\ref{pform}) and (\ref{newpform}).

While the ansatz, (\ref{eq:aantz}), may always be made, the localization
argument presented above indicates that at most only one of either $A_{[p]}$
or $A_{[p-1]}$ is bound to the brane---the former for $p<(d-2)/2$ and the
latter for $p>d/2$.  For intermediate values of $p$, there is no binding,
yielding the result that a $(d/2)$-form field strength on the brane cannot
originate from a bulk action of the form (\ref{bulkaction}) (but may arise
instead from odd-dimensional self-duality equations).  This is in contrast
to ordinary Kaluza-Klein reduction on a circle, where both potentials
survive in the massless sector.

Of course, (\ref{eq:eom1}) is not the only equation of motion we must
consider. There is also the bulk Einstein equation with cosmological
constant $\Lambda=-d(d-1)k^2$:
\begin{eqnarray}
{\hat R}_{MN}-\frac{1}{2}{\hat g}_{MN}\hat R&=&-\fft12{\hat g}_{MN}\Lambda
+\fft1{2\cdot p!}\Biggl[\hat g^{M_{1}N_{1}}\hat g^{M_{2}N_{2}}\cdots
\hat g^{M_{p}N_{p}}
{\hat F}_{MM_{1}M_{2}\ldots M_{p}}
{\hat F}_{NN_{1}N_{2}\ldots N_{p}}\nonumber\\
&&\kern-1em-\frac{1}{2(p+1)}{\hat g}_{MN}
\hat g^{M_{1}N_{1}}\hat g^{M_{2}N_{2}}\cdots \hat g^{M_{p+1}N_{p+1}}
{\hat F}_{M_{1}M_{2}\ldots M_{p+1}}
{\hat F}_{N_{1}N_{2}\ldots N_{p+1}}\Biggr].\nonumber\\
\end{eqnarray}
Substituting the ansatz (\ref{metric}) and (\ref{pform}) yields
\begin{eqnarray}
\label{eq:kk1eins}
{R}_{\mu\nu}-\frac{1}{2}{g}_{\mu\nu}R&=&\fft1{2\cdot p!}
e^{2pk|z|}\Biggl[g^{\mu_{1}\nu_{1}}g^{\mu_{2}\nu_{2}}\cdots g^{\mu_{p}\nu_{p}}
{F}_{\mu\mu_{1}\mu_{2}\ldots \mu_{p}}
{F}_{\nu\nu_{1}\nu_{2}\ldots \nu_{p}}\nonumber\\
&&-\frac{1}{2(p+1)}{g}_{\mu\nu}
g^{\mu_{1}\nu_{1}}g^{\mu_{2}\nu_{2}}\cdots g^{\mu_{p+1}\nu_{p+1}}
{F}_{\mu_{1}\mu_{2}\ldots \mu_{p+1}}
{F}_{\nu_{1}\nu_{2}\ldots \nu_{p+1}}\Biggr]
\end{eqnarray}
from the $\mu\nu$ components, and
\begin{equation}
\label{eq:kk1zz}
R=\fft1{2(p+1)!}e^{2pk|z|}
g^{\mu_{1}\nu_{1}}g^{\mu_{2}\nu_{2}}\cdots g^{\mu_{p+1}\nu_{p+1}}
{F}_{\mu_{1}\mu_{2}\ldots \mu_{p+1}}
{F}_{\nu_{1}\nu_{2}\ldots \nu_{p+1}}
\end{equation}
from the $zz$ component.
Since consistency of the Einstein equation requires that the
$z$-dependence cancel, this rules out all but $p=0$.  Alternatively, it is
straightforward to see by taking the trace of the first equation that
(\ref{eq:kk1eins}) and (\ref{eq:kk1zz}) are inconsistent unless $p=0$.

Similarly, substituting the ansatz (\ref{metric}) and (\ref{newpform}) yields
\begin{eqnarray}
{R}_{\mu\nu}-\frac{1}{2}{g}_{\mu\nu}R&=& \fft1{2(p-1)!}e^{2(d-p-1)k|z|}
\Biggl[g^{\mu_{1}\nu_{1}}g^{\mu_{2}\nu_{2}}\cdots g^{\mu_{p}\nu_{p}}
{F}_{\mu\mu_{1}\mu_{2}\ldots \mu_{p-1}}
{F}_{\nu\nu_{1}\nu_{2}\ldots \nu_{p-1}}\nonumber\\
&&\kern4em-\frac{1}{2p}{g}_{\mu\nu}
g^{\mu_{1}\nu_{1}}g^{\mu_{2}\nu_{2}}\cdots g^{\mu_{p}\nu_{p}}
{F}_{\mu_{1}\mu_{2}\ldots \mu_{p}}
{F}_{\nu_{1}\nu_{2}\ldots \nu_{p}}\Biggr]
\end{eqnarray}
and
\begin{equation}
R=-\fft1{2\cdot p!}e^{2(d-p-1)k|z|}
g^{\mu_{1}\nu_{1}}g^{\mu_{2}\nu_{2}}\cdots g^{\mu_{p}\nu_{p}}
{F}_{\mu_{1}\mu_{2}\ldots \mu_{p}}
{F}_{\nu_{1}\nu_{2}\ldots \nu_{p}},
\end{equation}
which rules out all but $p=(d-1)$. So when Kaluza-Klein consistency of the
coupled Einstein-$p$-form field equations is imposed, the only $p$-forms that
can be bound to the brane are the $0$-form and its (brane) dual the
$(d-2)$-form. This result could not have been found merely by
substituting the ansatz into the action and demanding convergence of the
$z$-integral. This highlights the well-known dangers of substituting a
Kaluza-Klein ansatz into an action rather than the equations of
motion \cite{Duffpope}. Note that in $d\leq4$, the Einstein equation
imposes no further restrictions. In particular in $d=4$, both $0$-forms
and $2$-forms are
still bound to the brane. Whereas in $d=6$, for example, only
$0$-forms and the dual $4$-forms are bound to the brane; the
$1$-form allowed by (\ref{converge}) and $3$-form allowed by
(\ref{newconverge}) are ruled out.

This also has interesting consequences for the phenomenon of ``screening''
discussed in \cite{kaloper}. There, the $p$-forms for which the $z$ integral
diverged were interpreted as valid $d$-dimensional modes, albeit modes whose
charge has been screened. However, we see that with the ansatze
(\ref{pform}) and (\ref{newpform}), there are no solutions of the coupled
Einstein $p$-form field equations except for $p=0$ and $p=(d-1)$, for
which the
integral converges. One may thus take one of two attitudes. If one is
content to take a test particle approach with no gravitational
dynamics, then the screening phenomenon may occur.  However, if one
takes the view that the combined gravitational and $p$-form dynamics
must be taken into account, then if screening occurs at all, it cannot
occur in the way described in \cite{kaloper} where the ansatz
(\ref{pform}) was employed.

\bigskip

We have enjoyed useful discussions with Hong L\"u, Eva Silverstein, and
Lisa Randall. We are especially indebted to Chris Pope for pointing
out the extra restrictions imposed by the Einstein equation. A somewhat
different approach to the duality problem has recently been presented in
\cite{Silverstein}.

\end{document}